\documentclass[twocolumn,superscriptaddress,showpacs,prl,amsmath,amssymb]{revtex4}         
\usepackage{graphicx}
\usepackage{bm}
\begin{document}         
         
\title{         
Two-step melting of the vortex solid in layered superconductors with         
random columnar pins}         
         
\author{ Chandan Dasgupta}         
\email{cdgupta@physics.iisc.ernet.in}         
\affiliation{         
Condensed Matter Theory Center,         
Department of Physics, University of Maryland, College Park, Maryland         
20742-4111}         
\affiliation{Department of Physics, Indian Institute of Science,         
Bangalore 560012, India}         
\author{ Oriol T. Valls}         
\email {otvalls@umn.edu}         
\affiliation{ Department of Physics         
and Minnesota Supercomputer Institute,         
University of Minnesota,         
Minneapolis, Minnesota 55455-0149}         
         
\begin{abstract}         
We consider the melting of the vortex solid in highly anisotropic         
layered superconductors with a small         
concentration of random columnar pinning centers.         
Using large-scale numerical minimization of a free-energy functional,         
we find that  melting of the low-temperature, nearly crystalline         
vortex solid (Bragg glass) into a vortex         
liquid occurs in two steps as the temperature          
increases: the Bragg glass and liquid phases are         
separated by an intermediate Bose glass phase.         
A suitably defined local melting         
temperature exhibits spatial variation similar to that observed         
in experiments.

\end{abstract}         
         
         
\maketitle         
         
The mixed phase of type-II superconductors with random pinning         
constitutes an excellent test system for studies of  the effects of quenched         
disorder on the structure and melting of crystalline solids. In         
systems with weak random point pinning, the existence of         
a low-temperature topologically         
ordered Bragg glass (BrG) phase with quasi-long-range translational order         
is now well established ~\cite{brg1,brg2}.         
A variety of fascinating ``glassy'' behavior has been experimentally         
observed~\cite{tifr}  near the first-order melting transition of the         
BrG phase in both conventional and high-$T_c$ superconductors. It has been         
suggested~\cite{tifr,menon} that these observations can be         
understood if it is assumed that the melting of the BrG phase occurs in two         
steps: the BrG first transforms into a ``multidomain'' glassy phase which         
melts into the usual vortex liquid at a slightly higher temperature.         
         
In the presence of random columnar pinning, a ``strong''         
Bose glass (BoG) phase~\cite{nelson} without         
quasi-long-range translational order occurs at low temperatures if the         
concentration of pins is larger than that of vortex lines. In the opposite         
limit of dilute pins, one expects~\cite{radz} a ``weak'' BoG phase at low         
temperatures which would melt         
into an interstitial liquid (IL) as the temperature is increased. In the IL         
phase, some of the vortices remain pinned at the strong pinning centers,         
while the other, interstitial ones form a liquid. A recent numerical         
study~\cite{bbrgl} suggests that a         
topologically ordered BrG phase         
is also possible in such systems if the pin concentration is         
sufficiently small. It is also         
found experimentally, for both point~\cite{expt1} and columnar~\cite{expt2}         
pinning, that the melting of the  solid phase is         
``broadened'': the local transition temperature, measured         
by a discontinuity of the local magnetization, is different in different         
regions of the sample.         
         
Here we report  results of a numerical study that provides         
insights and explanations for some of the  observations described above. From         
minimization of an appropriate free energy functional, we find         
that the vortex system in an extremely anisotropic, layered, superconductor         
with a random dilute array of strong columnar pins (with both pins and         
magnetic field         
perpendicular to the layers) forms a BrG phase at low         
temperatures. As $T$ is increased, this phase undergoes a          
first order transition into a glassy phase         
which we identify as a polycrystalline BoG. This phase then transforms,         
at a slightly higher $T$, into the IL phase via a second,         
more strongly first order transition.         
We also show that the local transition         
temperatures, obtained from the temperature-dependence of a quantity that         
measures the degree of localization of the vortices in a small region of the         
sample, exhibit substantial spatial variation  correlated with         
the local arrangement of the pinning centers.         
         
\begin{figure*}         
\includegraphics [scale=0.4]{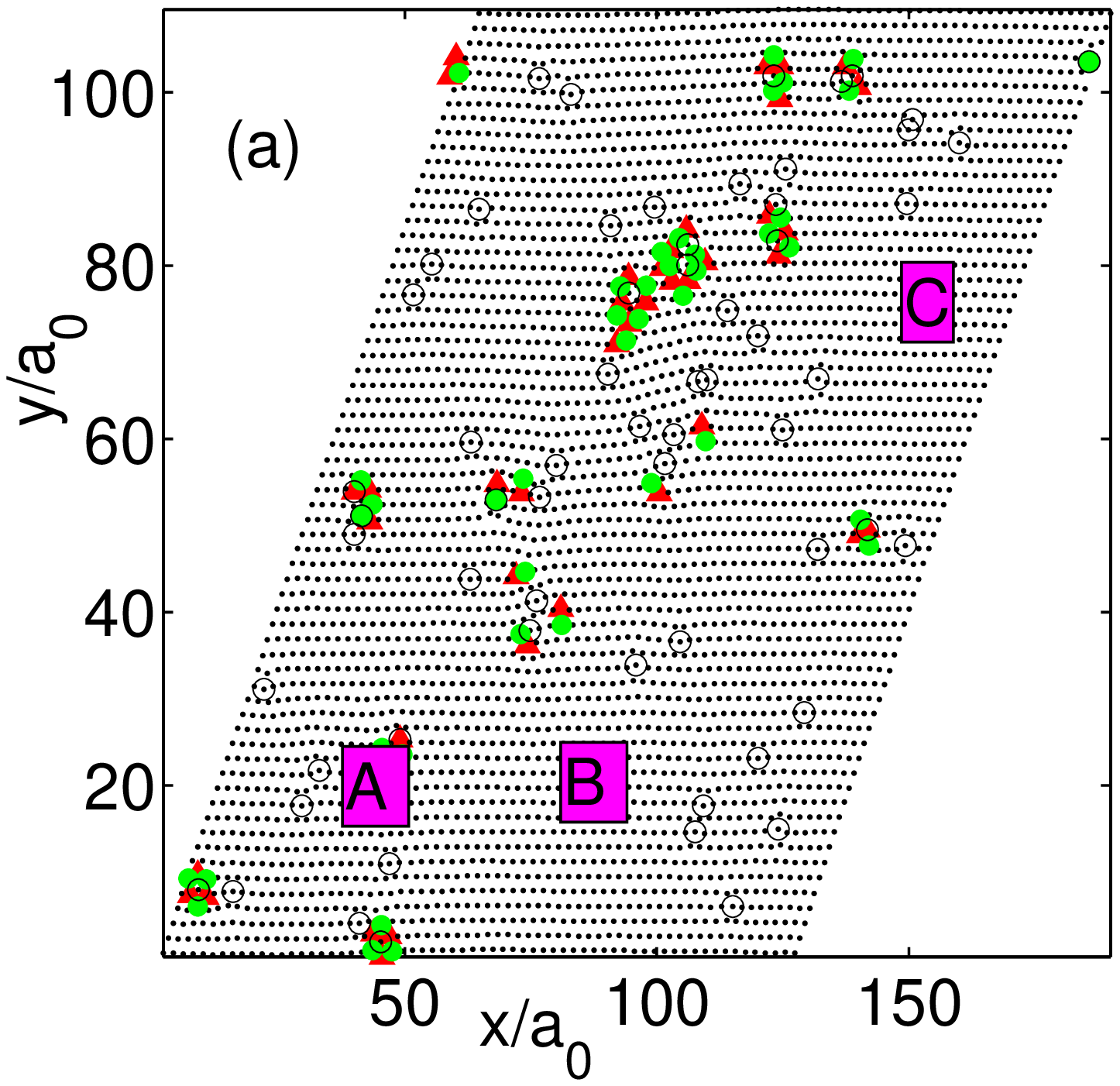}         
\includegraphics [scale=0.4]{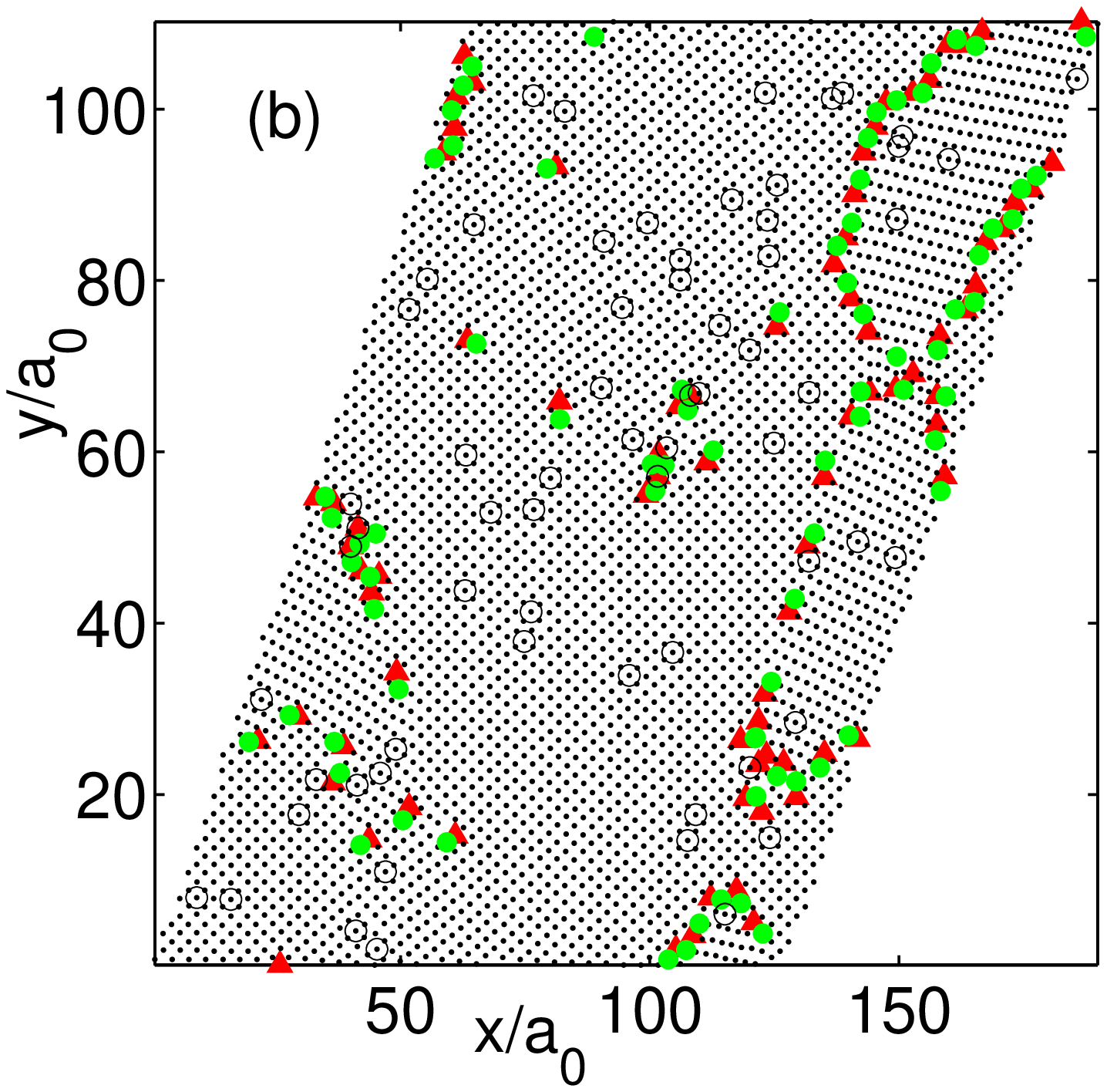}         
\includegraphics [scale=0.4]{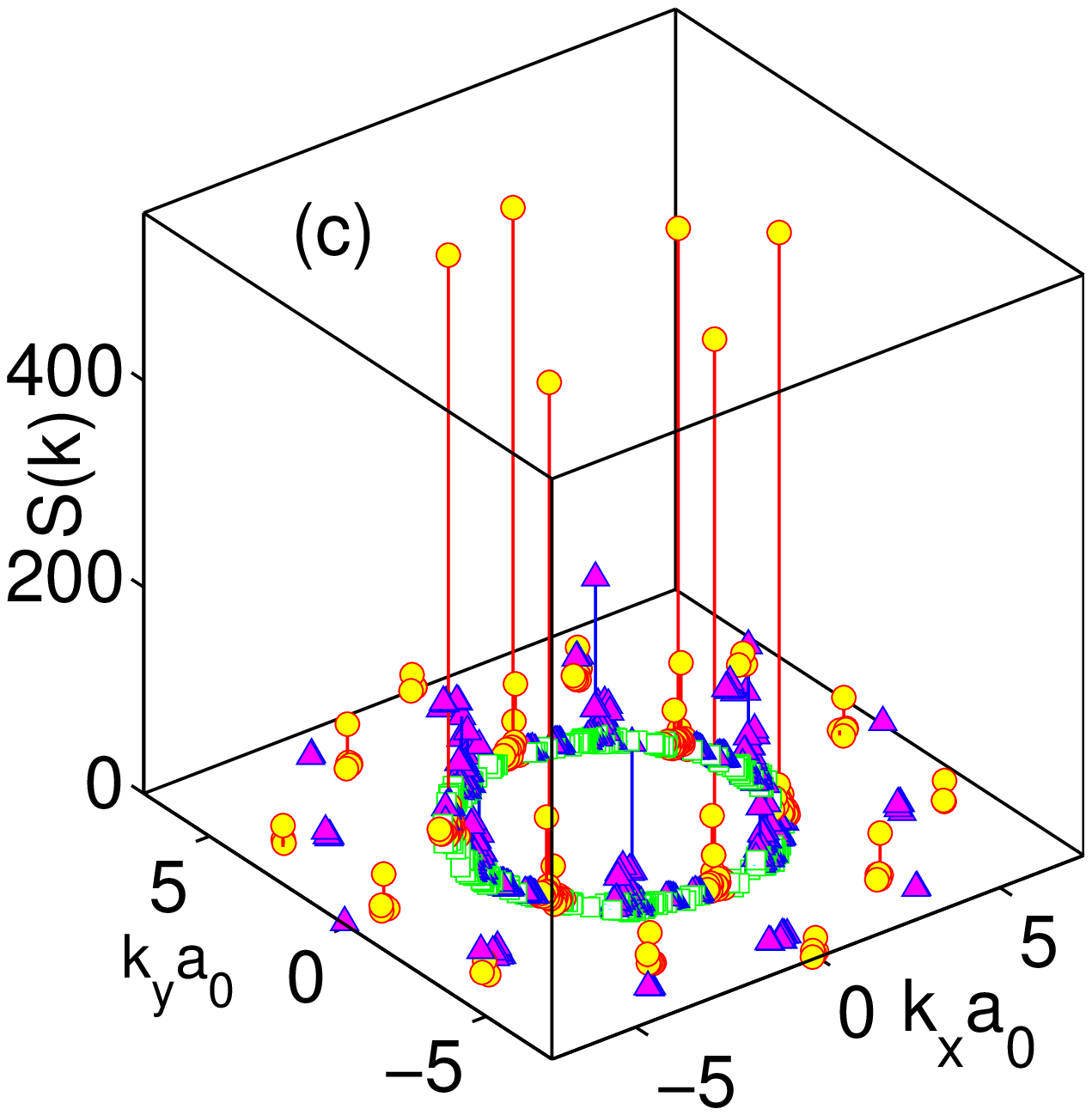}         
\caption{\label{fig1} (Color online.) Results for a sample with 4096         
vortices and 64 pins. Panel (a): Voronoi plot (see text) for the         
BrG minimum at 17.8K, with 5-, 6- and 7-         
coordinated sites indicated by (red) triangles, small (black) dots and         
large (green) dots, respectively. Black circles denote         
pin positions. Dislocations form tightly bound clusters near pin         
locations. Regions labeled A, B, C are discussed in the         
text and Fig.\ref{fig2}. Panel (b): Same as panel (a), but for a BoG minimum at         
18.2K. Here, dislocations are arranged in lines to form grain boundaries.         
Panel (c): Structure factors at 18.4K. Results for the BrG, BoG and IL         
minima are shown by (yellow) circles, (purple) triangles, and (green)       
squares, respectively. Vertical lines are guides to the eye.         
}         
\end{figure*}         
         
The model and  methods we use are similar         
to those in our earlier work~\cite{cdotv} for a periodic array of         
columnar pins: the main difference is that         
the pin array here is taken to be random. Thus,         
we study a layered superconductor with vanishingly small Josephson         
interlayer coupling (vortices on different layers are coupled via         
the electromagnetic interaction only). In this limit,          
appropriate~\cite{gautam} for extremely anisotropic Bi- and Tl-based         
high-$T_c$ materials, the energy of a system of ``pancake''         
vortices residing on the superconducting layers may be written as a sum         
of anisotropic two-body interactions. We use the         
Ramakrishnan-Yussouff (RY) free energy functional~\cite{ry}.         
Since the potential produced by a set of straight         
columnar pins perpendicular to the layers is the same on every layer,         
$\rho({\bf r})$, the {\it time-averaged} local areal density of vortices at         
point $\bf r$ on a layer,         
must be the same on all layers. The free energy $F$ per layer         
may then be written as:    
\begin{eqnarray}         
\beta (F[\rho] - F_0) &=& \int d^2r\left[\rho({\bf r})         
\{ \ln(\rho({\bf r}))-\ln (\rho_0)\} -         
\delta\rho({\bf r})\right] \nonumber \\         
&-&\frac{1}{2}\int d^2r \int d^2r^\prime \tilde{C}(|{\bf r} -         
{\bf r}^\prime |)         
\delta\rho({\bf r}) \delta\rho({\bf r}^\prime) \nonumber \\         
&+& \beta \int d^2r V_p({\bf r}) \delta\rho({\bf r}).         
\label{ryfe}         
\end{eqnarray}         
Here, $\delta\rho({\bf r}) \equiv \rho({\bf r}) - \rho_0$, $F_0$ is the free         
energy of the uniform liquid of areal density $\rho_0$ (= $B/\Phi_0$         
where $B$ is the magnetic induction and $\Phi_0$ the          
flux quantum), $\beta=1/k_BT\/$, $V_p({\bf r})$         
is the pinning potential, and $\tilde{C}(r) \equiv \sum_n C(n,r)$,         
where $C(n,r)$ is the {\it direct pair correlation function} of a         
layered liquid of pancake vortices ($n$ is the layer separation         
and $r$ is the separation in the layer plane).         
We use the results for $C(n,r)$         
obtained \cite{gautam} from a hypernetted chain calculation.         
The RY functional yields~\cite{cdotv,gautam} a correct {\it quantitative}         
description of the melting transition in the         
absence of pinning.         
The potential $V_0(r)$ at $\bf r$ due to a         
pinning center at the origin is assumed to have the form         
$V_0(r) = -\alpha \Gamma (1 - r^2/r_0^2)$         
for $r \leq r_0$ and $V_0(r) = 0$ if $r >r_0$. Here,         
$\Gamma \equiv \beta d \Phi^2_0/8 \pi^2 \lambda^2(T)\/$,         
$d$ is the layer spacing, $\lambda(T)$ is         
the penetration depth in the layer plane, $r_0$         
is a range parameter and $\alpha$ is a strength parameter. The net pinning         
potential $V_p({\bf r})$ is the sum of the potentials due to $N_p$ randomly         
placed pinning centers.         
We use parameter values appropriate to BSCCO i.e.  $\lambda(T=0) = 1500 \AA$         
and $d = 15 \AA$, and assume a two-fluid $T$-dependence of         
$\lambda(T)$ with $T_c(0)=85$K. Defining  $a_0$ (which we         
will use as our unit of length) via the         
relation $\pi a_0^2 \rho_0 =1$, we set $r_0=0.1 a_0$ and $\alpha=0.05$. For         
these values, each pinning center traps one         
vortex~\cite{cdotv} in the temperature range of interest.         
         
We discretize space by defining density variables $\{\rho_j\}$ at the sites of         
a triangular grid of size $(Nh)^2$ with periodic boundary conditions.         
The grid spacing $h$ is taken to be $a/16$ where $a\simeq 1.988a_0$ is the         
equilibrium spacing~\cite{cdotv} of the pure vortex lattice at         
melting for the value of the magnetic field ($B$ = 2kG) used here.         
The $N_p$ pinning centers are randomly put on computational lattice sites.         
Local minima of the discretized $F[\rho]$ Eq.(\ref{ryfe}), written as a         
function of  the $\{\rho_j\}$, are then obtained numerically using a         
methodology quite similar to that in Ref.~\cite{cdotv}.         
We report here primarily results for $N$ = 1024 (corresponding,         
for the chosen value of $h$, to including $N_v$ = 4096 vortices         
in the calculation), and relative  pin concentration (the ratio of         
the number of pins  $N_p$ to $N_v$) $c$ = 1/64. A         
larger pin concentration, $c$ =1/32, and samples of size $N$ = 512 (1024         
vortices) were also studied~\cite{prep}.         
Results for the glassy phases depend         
somewhat, see below, on the placement         
of the $N_p$ random pins.         
Over twenty different         
random pin configurations were studied and averages  taken where         
appropriate.         
         
\begin{figure}         
\includegraphics [scale=0.4]{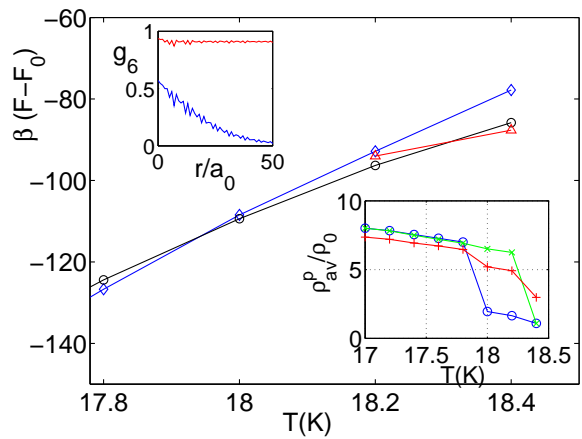}         
\caption{\label{fig2} (Color online). Main plot: Temperature dependence         
of the free energies of different minima of the sample of Fig.~\ref{fig1}.         
Results for the BrG, BoG and IL minima are shown by (blue) diamonds, (black)         
circles and (red) triangles, respectively. 
The crossings near 17.9K and 18.3K correspond to transitions.         
Upper inset: Bond-orientational correlation function $g_6(r)$ (see text),         
plotted vs. $r/a_0$, for BrG (upper (red) curve) and BoG (lower (blue)         
curve) minima at 17.0K and $c=1/64$. Lower inset: T-dependence         
of the average local peak density $\rho^p_{av}$ for small regions (see text)         
near the points A, B and C in Fig.\ref{fig1}(a). Results for A, B and C         
are shown by (red) plus signs, (green) crosses and (blue) circles,         
respectively. Solid lines are guides to the eye.}         
\end{figure}         
         
Different local minima of the discretized $F$, corresponding to         
the several possible phases of the system,         
are reached  when the numerical minimization is performed starting         
from different initial states~\cite{cdotv}. The free         
energies of these different minima at a given $T$ determine         
the (mean-field) phase diagram. A great advantage of our method         
in identifying minima         
representing different phases is that at each minimum         
the values of the  full $\{\rho_j\}$ set are available. We can then calculate         
e.g. the structure factor $S({\bf k})=|\rho({\bf k})|^2/N_v$.         
We can further characterize the structure of a minimum by analyzing additional         
information~\cite{prep}  contained in the full set $\{\rho_j\}$.         
In particular, we have  calculated the {\it local peak         
densities}, defined as the values of the density at {\it local} peaks.         
The  density is considered         
to locally peak at a mesh point $j$ if  $\rho_j$ 
is higher than those at all other mesh points within a distance $a/2$ from $j$.         
At low-temperature minima         
with localized vortices, these local density peaks lie         
where the vortices are localized: their number matches the         
number of vortices $N_v$. Thus, the positions of these  peaks         
define a ``vortex lattice''.         
To elucidate the degree of order in this ``vortex lattice'',         
we have found it particularly useful         
to carry out a Voronoi construction,         
thereby determining the number of nearest neighbors of each vortex and         
hence the defect structure of the minimum. To address questions         
about orientational order,         
we have obtained the bond-orientational         
{\it vortex} correlation function $g_6(r)$, defined as         
the correlation function of the field         
$\psi({\bf r})= \sum_j \exp[6i\theta_j({\bf r})]/n_n$, where         
$\theta_j({\bf r})$ is         
the angle that the bond connecting a vortex at $\bf r$ to its $j$th neighbor         
makes with a fixed reference direction, and $n_n$ is the number of neighbors         
of the vortex at $\bf r$.         
         
We find three different kinds of local minima of $F$.         
The simplest kind is found by quenching the system, starting from         
uniform initial conditions, to temperatures somewhat higher than the         
equilibrium melting temperature of the pure vortex lattice         
($T_m^0 \simeq 18.4$K~\cite{cdotv}).         
The minimum can then be slowly cooled to below $T_m^0$ until it         
becomes unstable. This phase         
has local vortex densities close to the uniform liquid density         
everywhere except near         
the pinning centers, each of which traps one vortex. These minima         
clearly correspond to the interstitial liquid (IL) state with very small         
($\le 5$) peak values of $S({\bf k})$         
(see panel (c) of Fig.\ref{fig1}).         
         
The second kind of local         
minima are  obtained by quenching to temperatures below $T_m^0$         
with initial conditions         
corresponding to a perfectly crystalline initial state (we         
use the crystalline state for which the pinning energy is minimum). They         
can then be cooled down, or warmed up to above $T_m^0$.         
This phase is nearly crystalline: the densities at local peaks         
turn out to be large (5-10 times $\rho_0$) nearly everywhere,         
except at the pins         
where they are much higher. There are also a few small regions         
of lower peak density, indicating weakly localized vortices.         
The Voronoi plots for such minima (see panel (a)         
in Fig.~\ref{fig1})         
clearly illustrate the defect structure. A pair of adjacent         
5- and 7-coordinated sites, shown as (red) triangles and         
large (green) dots respectively, corresponds         
to a dislocation. These dislocations form tightly bound clusters in         
this case. These clusters are located near pinning sites, shown by         
black circles in panels (a) and (b) of Fig.\ref{fig1}. 
The local peak densities near the defect         
clusters are lower, indicating weaker localization of the vortices.         
The  structure factor plot, shown in panel (c), exhibits six sharp         
Bragg peaks for these minima.         
The ``crystalline order parameter'' extracted from the peak value of          
$S({\bf k})$ is large at low temperatures ($\simeq 0.55$ at 17K for $N_v=1024$)         
and decreases very slowly with sample size (by $\sim$ 6\% as the sample size is         
doubled).          
The vortex bond-orientational correlation         
function $g_6(r)$ for such minima (see Fig.~\ref{fig2}, upper inset)         
saturates to a large value for large $r$. Thus, these minima exhibit all the         
characteristics~\cite{brg1} of a BrG phase and we conclude         
that they can be so identified. However, our numerical study can not rule         
out the occurrence of unpaired dislocations at much longer length scales.         
If this happens, then these minima may correspond to a ``hexatic         
glass''~\cite{hexatic} phase. In any case, it is clear that these minima       
represent a phase         
that is distinct from the polycrystalline one described below.         
         
The third kind of minima         
are obtained either by slowly cooling a liquid-like minimum to below         
the temperature where it becomes unstable, or by quenching with a uniform         
initial density to a temperature well below $T_m^0$.         
The Voronoi construction results for this case (panel (b) of Fig.~\ref{fig1})         
clearly show a polycrystalline structure         
with the dislocations lining up along grain boundaries, which         
lie mainly in regions without any pinning center. As a liquid-like initial         
state is cooled from relatively high $T$, the vortices arrange         
themselves in triangular crystalline patches around the pins. The         
orientation of a crystalline patch depends on the local pin arrangement.         
As the temperature is lowered further, misaligned  patches join         
each other at grain boundaries to form a new minimum         
of this kind. The local peak density         
is substantially lower near the grain boundaries. As these minima are warmed         
up\cite{prep}, the regions near the grain boundaries begin to         
``melt'' before the other         
parts of the sample.         
As shown in panel (c) of Fig.\ref{fig1}, the structure factor for these         
minima exhibits several (typically more than six) peaks of height much lower         
than that of the six Bragg peaks found for the BrG minimum. 
The function $g_6(r)$ for such minima         
goes to zero at large $r$ (see Fig.~\ref{fig2}, upper inset).         
We conclude, therefore,         
by considering all the evidence, that such minima correspond         
to polycrystalline BoG states.  While for any given pin configuration the         
IL and BrG states reached upon the         
minimization are, within numerical uncertainty, unique, different         
BoG type states can be reached using different starting states and         
cooling rates. This is characteristic of a         
glassy phase. When more than one BoG minima         
are found at a given $T$, we consider the one with the         
lowest free energy. The polycrystalline         
nature of the BoG minima is consistent with the results of         
experiments~\cite{expt3} and simulations~\cite{nandini} of the Bose         
glass phase in the dilute-pin regime.         
         
Fig.~\ref{fig2} illustrates the main result of our study.         
There we show the temperature dependence of         
$\beta (F-F_0)$ of the BrG, IL  and BoG minima for the         
same pin configuration at $c$=1/64.         
The BrG minimum has the lowest free energy at         
low temperatures. Its free energy crosses that of the BoG minimum near 17.9K,         
indicating a first-order BrG-BoG transition at this temperature. The BoG phase         
then melts into the IL phase at a slightly higher temperature, near 18.3K, as         
indicated by the crossing of the free energies of these two minima. Thus, the         
melting of the low-temperature BrG phase with increasing temperature occurs         
in two steps, with a small region of BoG phase separating the BrG and IL         
phases. The grain boundaries in the BoG minima survive thermal          
cycling~\cite{expt3} across the BoG-IL transition temperature, whereas         
similar thermal cycling produces grain boundaries in the BrG minima,          
indicating that the BoG phase is thermodynamically stable           
near the BoG-IL transition.         
         
For samples with $c=1/64$,         
the average value of the temperature interval in which the BoG         
is the equilibrium state is 0.42 K. This width varies         
from sample to sample between less than 0.1 in one         
single case, to 1.2K.         
Thus, the two-step transition is a generic feature.         
The entropy jump at the BrG-BoG (lower) transition ($\simeq 0.1k_B$ per          
vortex) is smaller than that at the BoG-IL transition ($\simeq 0.15k_B$ per         
vortex). These values do not depend significantly on the system         
size and their sum is slightly smaller than      
the value ($0.29 k_B$) at the single melting transition in      
the pure system. The size of these jumps makes it very unlikely that    
fluctuations would   
change the nature of the transitions in our 3D system.    
The value of the upper transition temperature is       
between 18.2K and 18.3K for all samples at $c$ = 1/64. These values are       
close to the first-order melting temperature of the pure          
system~\cite{cdotv}.         
The weak dependence  of the transition temperatures on $c$ is consistent with         
experiments~\cite{expt2,expt3}. Also, a first-order freezing transition of         
the IL to a polycrystalline solid has been observed~\cite{expt3}         
in experiments on BSCCO with a small concentration of columnar pins.         
A narrow ``two-phase'' region found near the BrG melting transition in          
Ref.\cite{bbrgl} may correspond to an intermediate BoG         
phase. Alternatively, the sample size ($\sim 100$ vortex lines) 
in Ref.~\cite{bbrgl} may be too small
for the detection of an intermediate BoG phase with large crystalline domains.         
   
The BrG-BoG transition occurs as a result of a competition         
between elastic and pinning parts of the free energy.         
The BrG minimum has a lower elastic         
(free) energy than the BoG minimum, but a higher pinning energy:         
the vortices adjust better to the pinning potential in the BoG minimum.         
The softening of the lattice near  melting decreases the         
relative 
importance of the elastic component, thus causing a crossing of         
the two free energies.         
         
We have also calculated a space-dependent ``local transition temperature''         
by monitoring the temperature-dependence of $\rho^p_{av}$, the average of         
the local peak density in small regions containing $\sim$ 100 vortices.         
Vortices localized at pinning centers are {\it not} included in the         
calculation of $\rho^p_{av}$.         
Values of $\rho^p_{av}$ much larger than $\rho_0$ indicate a solid-like         
local structure, while values close to $\rho_0$ indicate         
liquid-like behavior. In Fig.\ref{fig2}, lower inset, we show the         
$T$-dependence of  $\rho^p_{av}$ for three regions centered at points         
A, B and C in panel (a) of         
Fig.\ref{fig1}. The local transition temperature,         
defined as         
the temperature at which $\rho^p_{av}$ crosses $3 \rho_0$~\cite{cdotv},         
is different in the three         
regions. The lowest local $T_c$ corresponds to the BrG-BoG         
transition and reflects the local melting near a grain boundary of the BoG         
minimum (point C). The highest local $T_c$ is higher than the BoG-IL         
transition temperature, reflecting solid-like local structure near a cluster         
of pinning sites in the IL minimum (point A). The range of variation of the         
local $T_c$'s is comparable to that found in experiments~\cite{expt1,expt2}.         
         
Thus, we have shown that a layered superconductor with a small         
concentration of columnar pins exhibits a two-step melting transition         
from a low-$T$ BrG phase to a high-$T$ IL phase via an intermediate BoG         
phase. A suitably defined local transition temperature         
exhibits spatial variations correlated with the local arrangement of pinning         
centers.         
Our results are consistent with  experiment, and         
support the suggestion~\cite{tifr,menon} of similar behavior         
in systems with point pinning.

\end{document}